\documentclass{article}

\usepackage{arxiv}

\usepackage[utf8]{inputenc} % allow utf-8 input
\usepackage[T1]{fontenc}    % use 8-bit T1 fonts
\usepackage{hyperref}       % hyperlinks
\usepackage{url}            % simple URL typesetting
\usepackage{booktabs}       % professional-quality tables
\usepackage{amsfonts}       % blackboard math symbols
\usepackage{nicefrac}       % compact symbols for 1/2, etc.
\usepackage{microtype}      % microtypography
\usepackage{graphicx}
\usepackage{amsmath}
\usepackage{natbib}
\usepackage{listings}
\usepackage{xcolor}

\title{Open Wave Logger v2026 (OWL-v2026): an open source, low cost, easy to build, high performance logger for wave data measurements}

\author{
  Jean Rabault \\
  Norwegian Meteorological Institute \\
  Oslo, Norway \\
  \texttt{jean.rblt@proton.me} \\
  \And
  Joey Voermans \\
  The University of Melbourne \\
  Melbourne, Australia \\
  \And
  Takuji Waseda \\
  The University of Tokyo \\
  Kashiwa, Japan \\
  \And
  Takehiko Nose \\
  The University of Tokyo \\
  Kashiwa, Japan \\
  \And
  Tsubasa Kodaira \\
  The University of Tokyo \\
  Kashiwa, Japan \\
  \And
  Koya Sato \\
  The University of Tokyo \\
  Kashiwa, Japan \\
  \And
  Alexander Babanin \\
  The University of Melbourne \\
  Melbourne, Australia \\
  \And
  Gaute Hope \\
  Norwegian Meteorological Institute \\
  Bergen, Norway \\
  \And
  Malte M\"{u}ller \\
  Norwegian Meteorological Institute and University of Oslo \\
  Oslo, Norway \\
  \And
  Lars Willas Dreyer \\
  University of Oslo \\
  Oslo, Norway \\
  \And
  \O{}ystein Lande \\
  University of Oslo \\
  Oslo, Norway \\
  \And
  Atle Jensen \\
  University of Oslo \\
  Oslo, Norway \\
  \And
  \O{}yvind Breivik \\
  Norwegian Meteorological Institute and University of Bergen \\
  Bergen, Norway \\
}

\begin{document}

\maketitle

\begin{abstract}
Ocean wave models are critical for weather and climate forecasting, and accurate in-situ wave observations are essential for validating and improving these models. Open-source, community-driven buoys have democratized wave observations via telemetry in recent years, but these systems transmit only limited amounts of data. Full high-frequency time series, required to study detailed wave physics, can still in most cases only be collected in situ using data loggers. Yet open-source, low-cost logger solutions remain scarce compared to their telemetry-enabled counterparts. Here we present the Openlogartemis Wave Logger (OWL-v2026), an open-source, low-cost, easy-to-build, high-performance logger for wave data measurements. The OWL-v2026 is built from off-the-shelf components from the maker community, requiring only through-hole soldering for assembly, and totals approximately 220~USD per unit. Custom firmware enables high-frequency, low-jitter logging of six-axis inertial measurement unit (IMU) data at 208 or 416~Hz, and GNSS position and Doppler velocity at 10~Hz, with Pulse Per Second (PPS) synchronization for accurate absolute UTC timestamping. We have successfully validated continuous logging over more than 10 days at 208~Hz, a power consumption of approximately 80~mA (approximately 20 days of autonomy with three D-cell lithium batteries), and absolute UTC timestamp accuracy typically better than 10~ms. Though the OWL-v2026 is a purely technical contribution, it has the potential to substantially expand the availability and affordability of high-frequency in-situ wave time series, similar to how the OpenMetBuoy (OMB) \citep{rabault2022openmetbuoy} expanded the availability of telemetry-enabled wave observations and helped spark new developments in low-cost open-source buoys.
\end{abstract}

\section{Introduction}
\label{sec:intro}

Ocean wave models are central to weather and climate forecasting, with applications spanning maritime safety, coastal management, and offshore energy production \citep{cavaleri2007wave,janssen1996ocean,meucci2020projected}. State-of-the-art operational wave models perform well in open-ocean conditions but have known limitations in a number of conditions where additional complexity arises, such as regions with complex bathymetry, coastal environments, and sea ice cover \citep{halsne2022resolving,bhaskaran2019challenges,shen2019modelling,collins2025measuring}. Several wave phenomena also remain incompletely understood, including wave breaking, freak waves, sea ice drift under complex forcing, and waves propagating in sea ice, and progress on these requires detailed observational data \citep{babanin2011breaking,didenkulova2006freak,sutherland2022estimating,shen2022wave,nose2023comparison,Cheng2025attenuation}. In sea ice in particular, recent observational studies have documented effects such as ice floe collisions \citep{dreyer2025direct}, strongly modulated wave attenuation \citep{rabault2024situ}, and nonlinear wave triads \citep{rabault2025observations}. Understanding these phenomena quantitatively requires access to high-resolution full time series, ideally from multiple sensors within a domain spanning several wavelengths. In addition, gathering more in-situ data is critical to calibrating parameterizations for complex phenomena such as waves in ice, for which models are still imperfect \citep{yu2022,nose2024observation,kousalECMWFWavesInIce,rogers2025ice}.

Wave observations rely on both commercial and open-source buoys. Commercial instruments have been deployed for many decades, though traditionally at high cost \citep{krogstad1999some}; open-source community-driven alternatives and modern small-size low-cost commercial buoys \citep{kohout2015device, thomson2019new, raghukumar2019directional, waseda2017arctic, rabault2022openmetbuoy, thomson2023development, kodaira2023affordable, cavaleri2025more,rabault2023dataset, yurovsky2026very, dreyer2026olb} have emerged in recent years and substantially lowered the barrier to entry, seeing increasing deployment and scientific impact.

Generally these buoys rely on telemetry for data transmission, which imposes a fundamental constraint: due to power and cost limitations, telemetry links can carry only limited amounts of data per unit time. When the application demands high-frequency continuous time series at 100s to 1000s of Hz — as required for detailed wave physics — there is no globally connected alternative to a data logger that stores data locally and is retrieved in the field after deployment \citep{rabault2017measurements, YUROVSKY2020108043}. Despite the proliferation of open-source telemetry-enabled buoys, affordable open-source data loggers for wave measurements remain comparatively scarce. Existing open-source logger designs either predate current low-cost high-performance hardware and are consequently expensive and power-hungry \citep{rabault2017measurements}, or use firmware not optimized for low-jitter, high-frequency logging \citep{lande2026feasibility}. There is thus a clear need for an affordable, reliable, easy-to-build, high-performance open-source wave logger.

Here, we present the Openlogartemis Wave Logger (OWL-v2026) to fill this gap. The OWL-v2026 uses off-the-shelf breakout boards from the maker community that integrate industrial-grade sensors; it requires only through-hole soldering for assembly, and all code, schematics, and assembly instructions are publicly available. The key differentiator from prior open-source logger designs is the firmware: written from the ground up for this application, it delivers reliable, low-jitter wave motion logging at 208 or 416~Hz. The overall design is intentionally simple enough that groups without specialized electronics expertise can build and deploy OWL-v2026 units at scale — similar in spirit to e.g. the OpenMetBuoy (OMB) \citep{rabault2022openmetbuoy} and the Small FriendlY buoy (SFY) \citep{hope2025sfy}.

The low cost and ease of assembly also open new possibilities for large-scale distributed measurements. Arrays of tens to hundreds of OWL-v2026 units could enable, for example, direct cross-correlation analysis of wave fields, extending significantly previous wave in ice measurements \citep{rabault2025observations}, or two-dimensional spatial monitoring of wave dynamics over areas spanning several wavelengths. For typical swell propagating in sea ice at 150~m wavelength, a 50k~USD project could realistically deploy over 250 OWL-v2026 units, covering a 5-by-5 wavelength area with 50~m sensor spacing — a scale of spatial coverage previously achievable only in laboratory wave tanks.

\section{Design}
\label{sec:design}

\subsection{Design philosophy and components}
\label{subsec:components}

The OWL-v2026 is built around off-the-shelf, well-documented, affordable breakout boards from the maker community. This choice eliminates the most demanding aspects of custom electronics design — RF trace layout and impedance matching — while keeping component sourcing simple. All inter-board connections use low-frequency I$^{2}$C (Qwiic) and SPI interfaces, which impose no high-frequency trace requirements nor electrical engineering expertise on the scientist doing the assembly. The resulting hardware assembly process requires only hand-soldering of through-hole components at standard 2.54~mm pitch, making assembly feasible in any lab without specialized equipment.

Table~\ref{tab:components} lists the main components of the OWL-v2026, their functions, and their prices as of February 2026. The total cost per unit is approximately 220~USD. This represents a roughly ten-fold cost reduction compared to previous high-performance wave loggers based on commercial IMU systems such as the Vectornav VN100 \citep{rabault2017measurements}.

\begin{table}[htbp]
    \centering
    \caption{OWL-v2026 components and costs (prices checked February 2026). Purchase links are listed in the GitHub repository (Appendix~\ref{app:release}).}
    \label{tab:components}
    \begin{tabular}{llr}
        \toprule
        Component & Function & Price (USD) \\
        \midrule
        Pololu S7V8F3 & Step-up/step-down 3.3~V regulator & 9.95 \\
        OpenLog Artemis (no IMU) & Main logger board & 48.95 \\
        ISM330DHCX breakout & 6-axis IMU & 25.50 \\
        SAM-M10Q breakout & GNSS module with patch antenna & 51.50 \\
        SD card (32~GB Class~10) & Data storage & 32.95 \\
        Hammond 1554G2GY (IP68) & Waterproof enclosure & 30.00 \\
        D-cell battery holder & Battery holder & 5.00 \\
        Custom PCB & Assembly and power distribution & 5.00 \\
        Small extras (screws, switch, etc.) & Assembly hardware & 10.00 \\
        \midrule
        \textbf{Total} & & \textbf{218.85} \\
        \bottomrule
    \end{tabular}
\end{table}

\subsection{Data acquired}
\label{subsec:data}

The OWL-v2026 logs four complementary data streams, summarized in Table~\ref{tab:data}. The high-rate IMU data are the primary wave measurement channel, providing six-axis inertial data at 208 or 416~Hz. The GNSS provides position, altitude, and Doppler velocity in the North-East-Down (NED) frame at 10~Hz, enabling trajectory tracking and an independent velocity estimate. The Pulse Per Second (PPS) signal marks the start of each UTC second with microsecond-level precision and enables the absolute UTC timestamp calibration described in Section~\ref{subsec:decoder}. Every data frame is additionally tagged with the MCU microsecond counter, providing the raw timing that the decoder maps to UTC via the PPS regression.

\begin{table}[htbp]
    \centering
    \caption{Summary of data acquired and logged by the OWL-v2026.}
    \label{tab:data}
    \begin{tabular}{lllp{6.5cm}}
        \toprule
        Source & Quantity & Rate & Notes \\
        \midrule
        MCU & Microsecond timestamp & Per measurement & MCU $\mu$s since boot, attached to each data frame \\
        SAM-M10Q & GNSS fix & 10~Hz & Latitude, longitude, altitude, and Doppler velocity (NED) \\
        SAM-M10Q & PPS & 1~Hz & Rising edge of each UTC second, logged via interrupt \\
        ISM330DHCX & Accelerometer + gyroscope & 208 or 416~Hz & 3-axis acceleration and 3-axis angular rate in the sensor frame \\
        \bottomrule
    \end{tabular}
\end{table}

\subsection{Physical assembly}
\label{subsec:assembly}

\begin{figure}[htbp]
    \centering
    \includegraphics[width=0.65\linewidth]{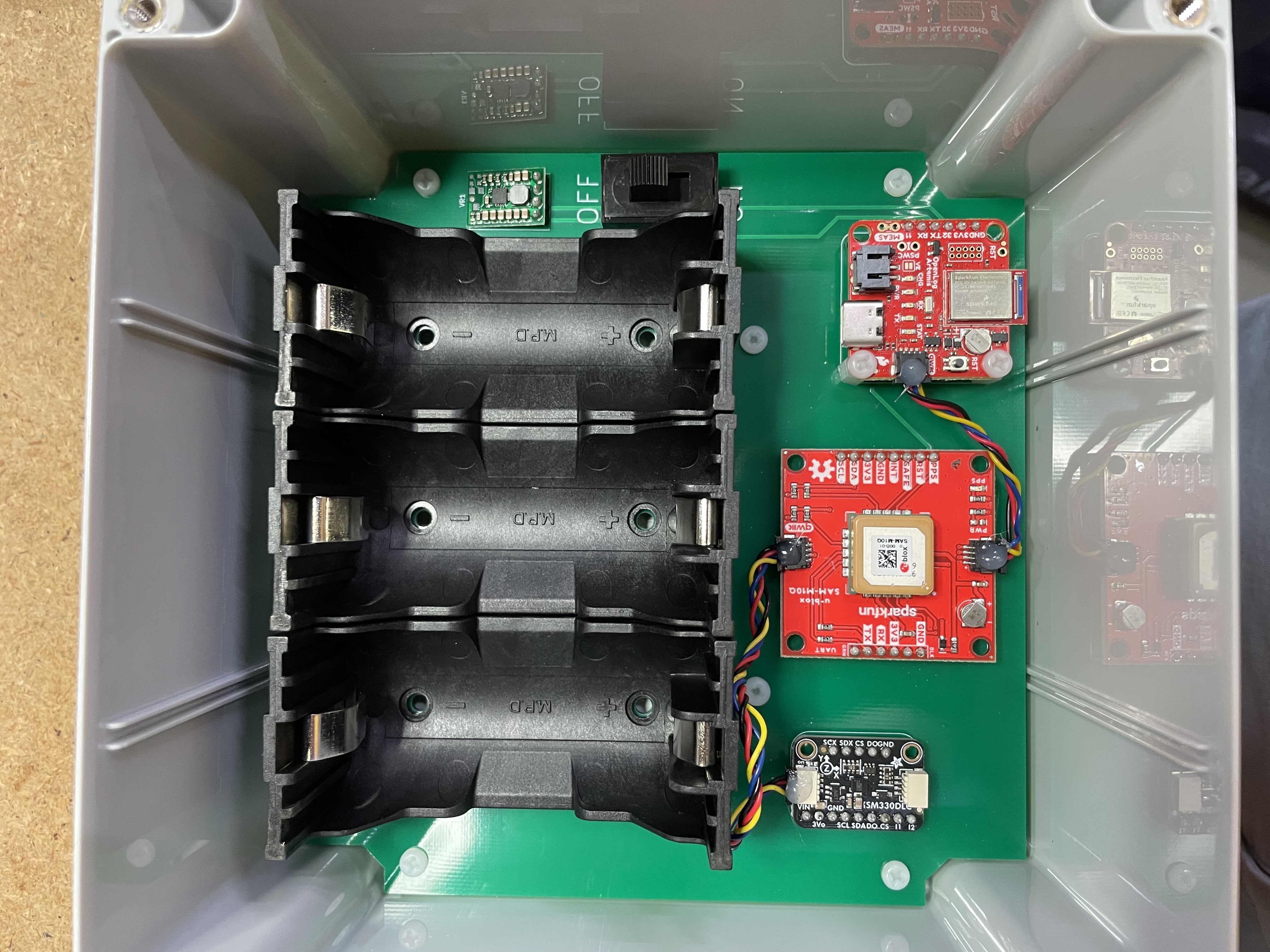}
    \caption{Assembled OWL-v2026 inside its IP68 enclosure, following the assembly instructions in the GitHub repository. Left: three D-cell battery holders connected in parallel (for use with lithium batteries such as SAFT LSH20). Top left: Pololu step-up/step-down power regulator and on/off switch. Top right: OpenLog Artemis main logger board. Right center: SAM-M10Q GNSS module. Bottom right: ISM330DHCX IMU breakout. Hot glue secures connectors to prevent disconnections in vibration-rich environments. Nylon screws fix the PCB to the enclosure.}
    \label{fig:pcb}
\end{figure}

All components are housed in a Hammond IP68 waterproof enclosure. A custom PCB organizes the breakout boards on a single plane, routes power from three D-cell lithium batteries (e.g., SAFT LSH20, connected in parallel) through the step-up/step-down regulator, and hosts the on/off switch. This makes switching the OWL logger ON/OFF and changing batteries easy in the field. Because the PCB uses only through-hole components at standard 2.54~mm pitch, assembly requires nothing more than a soldering iron. Additional robustness measures include hot glue on all Qwiic connectors to prevent vibration-induced disconnection during transport and field operation, and nylon screws to fix the PCB to the enclosure without risk of electrostatic discharge (ESD) damage. A fully assembled OWL-v2026 is shown in Figure~\ref{fig:pcb}.

\subsection{Firmware}
\label{subsec:firmware}

The firmware is the critical design element of the OWL-v2026. While generally similar hardware configurations have been used in previous wave loggers \citep{lande2026feasibility}, the OWL-v2026 firmware is designed from the ground up for reliable, low-jitter, high-frequency logging. The key design choices are:

\begin{itemize}
    \item \textbf{Bare-metal execution.} The firmware runs directly on the hardware without a real-time operating system (RTOS). This eliminates RTOS scheduling overhead and jitter, both incompatible with the low-latency, deterministic timing required for high-frequency sensor acquisition.
    \item \textbf{Interrupt-driven data acquisition.} Both IMU and GNSS data are acquired in hardware timer (ctimer) interrupt service routines (ISRs). This ensures low and consistent timing jitter, independent of main loop activity.
    \item \textbf{Multi-level buffering.} The ISM330DHCX provides a hardware FIFO. Data read from the ISM330DHCX FIFO and the GNSS in the ISR are placed into MCU RAM ring buffers sized to hold 20~seconds of data. The main loop asynchronously drains these buffers to the SD card, fully decoupled from the ISR. This architecture tolerates the variable write latency of SD cards without data loss.
    \item \textbf{Critical sections.} Shared data structures between the ISR and the main loop are protected using critical sections to prevent race conditions.
    \item \textbf{Watchdog timer.} A hardware watchdog timer resets the MCU on any software or hardware malfunction, enabling reliable long-term unattended operation.
    \item \textbf{Fault-tolerant startup.} At boot, the firmware verifies all components are functional and acquires a valid UTC time fix before logging begins. Any failure triggers a watchdog reset and retry.
\end{itemize}

\begin{figure}[htbp]
    \centering
    \includegraphics[width=0.90\linewidth]{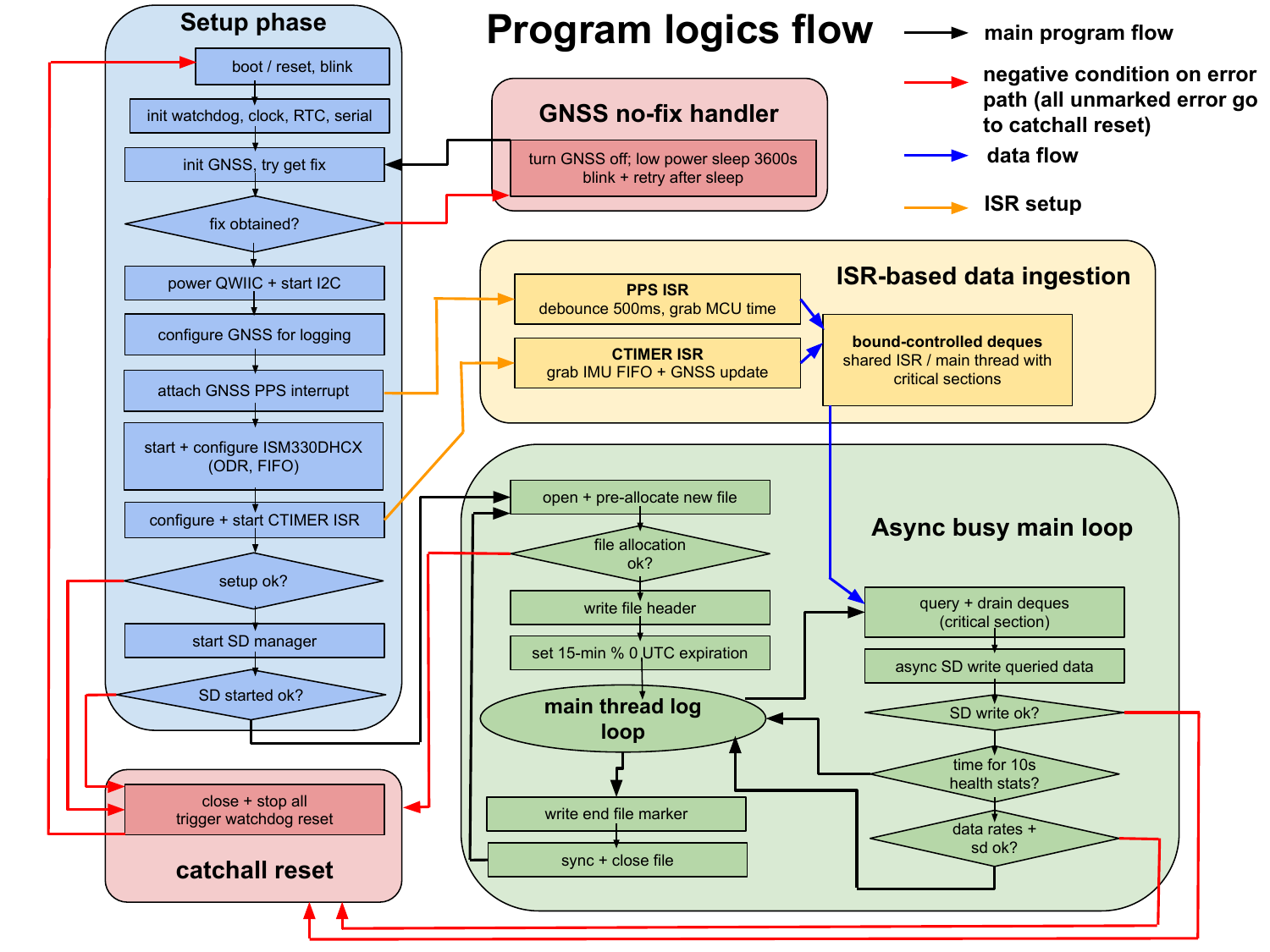}
    \caption{OWL-v2026 firmware program flow. At boot, the setup phase configures the GNSS, IMU, SD card, and MCU, verifies that all components are functional, and acquires a UTC time fix. Once setup is complete, logging runs as a collaboration between the asynchronous main loop (SD card writes) and the ctimer ISR (sensor data acquisition). A hardware watchdog timer resets the logger and forces a reboot on any malfunction. The firmware opens a new data file every 15 minutes.}
    \label{fig:flow}
\end{figure}

\begin{figure}[htbp]
    \centering
    \includegraphics[width=0.82\linewidth]{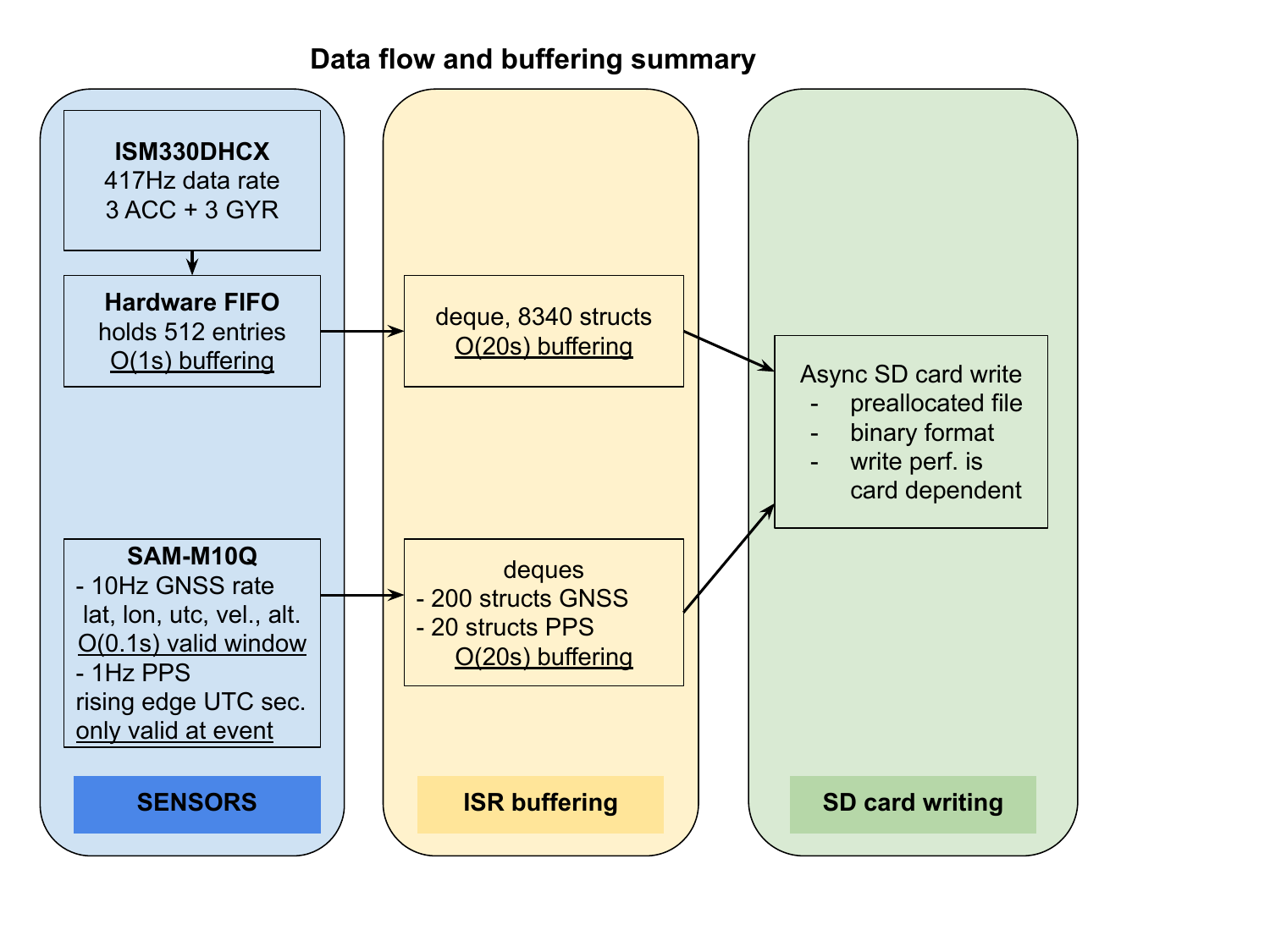}
    \caption{OWL-v2026 data flow from sensors to SD card. The ISM330DHCX hardware FIFO and the SAM-M10Q last-fix register act as sensor-side buffers. The ctimer ISR drains these and writes to MCU RAM ring buffers sized for 20~seconds of data. The main loop asynchronously writes the RAM buffers to the SD card. This multi-level architecture ensures reliable data acquisition despite the variable write latency of SD cards.}
    \label{fig:databuffer}
\end{figure}

The overall program flow is illustrated in Figure~\ref{fig:flow} and the data flow through the buffering stages in Figure~\ref{fig:databuffer}. Together, these design choices make the firmware both high-performance and robust: bare-metal execution and interrupt-driven acquisition minimize jitter; multi-level buffering decouples the deterministic sensor sampling from the intrinsically variable SD card write latency; and the watchdog combined with fault-tolerant startup ensure reliable operation over extended unattended deployments.

\subsection{Data storage and decoding}
\label{subsec:decoder}

Data are written to the SD card in a compact binary format for efficiency. Each frame on disk consists of a 4-byte ASCII tag (\texttt{\textbackslash nPPS}, \texttt{\textbackslash nGPS}, or \texttt{\textbackslash nIMU}) followed by the raw bytes of the corresponding C++ struct, shown in Listing~\ref{lst:structs}. This layout is self-documenting: the on-disk format is fully determined by the struct definitions in the firmware source code. To limit the impact of potential data corruption and keep file sizes manageable, the firmware opens a new file every 15~minutes (when the UTC second count is divisible by~900). The firmware pre-allocates each new file to a fixed size at the start of each logging segment to reduce SD card write latency at file boundaries.

\begin{lstlisting}[float=htbp, caption={Binary frame format written to the SD card. Each frame is identified by a 4-byte ASCII tag, followed by the raw bytes of the corresponding C++ struct. The struct definitions in the firmware source code fully specify the on-disk layout, making the format self-documenting.}, label=lst:structs]
// Each frame: 4-byte tag + struct (written as raw bytes)
//   "\nPPS" + PPS_fix  |  "\nGPS" + GNSS_reading  |  "\nIMU" + IMU_reading

struct PPS_fix {
  uint32_t micros_reading; // MCU microseconds at PPS rising edge
};

struct GNSS_reading {
  uint32_t micros_reading; // MCU microseconds at GNSS fix
  int32_t  latitude;            // degrees x 1e-7
  int32_t  longitude;           // degrees x 1e-7
  uint32_t posix_timestamp;     // POSIX UTC seconds
  uint32_t microseconds;        // sub-second microseconds
  int32_t  NED_vel_north;       // mm/s
  int32_t  NED_vel_east;        // mm/s
  int32_t  NED_vel_down;        // mm/s
  uint8_t  fix_type;            // 0 = no fix, 3 = 3D fix
};

struct IMU_reading {
  uint32_t micros_reading; // MCU microseconds at sample
  uint16_t counter;             // ISR sample counter
  int16_t  acc_x;               // raw ADC; 0.061 mg/LSB (+/-2 g full scale)
  int16_t  acc_y;
  int16_t  acc_z;
  int16_t  gyr_x;               // raw ADC; 4.375 mdps/LSB (+/-125 dps full scale)
  int16_t  gyr_y;
  int16_t  gyr_z;
};
\end{lstlisting}

A Python decoder converts the SD card binary data to user-friendly NumPy arrays. The decoder performs the following steps:
\begin{enumerate}
    \item Extract all binary frames from each SD card file and convert them to NumPy arrays.
    \item Stitch IMU, GNSS, and PPS data streams together and check for gaps.
    \item Split data into 1-minute segments. On each segment, fit a linear regression from the MCU microsecond timestamp to the UTC PPS timestamps, and apply this regression to produce UTC-timestamped IMU and GNSS data. Wrapping of the uint32 MCU microsecond counter is handled automatically.
\end{enumerate}

\begin{figure}[htbp]
    \centering
    \includegraphics[width=0.95\linewidth]{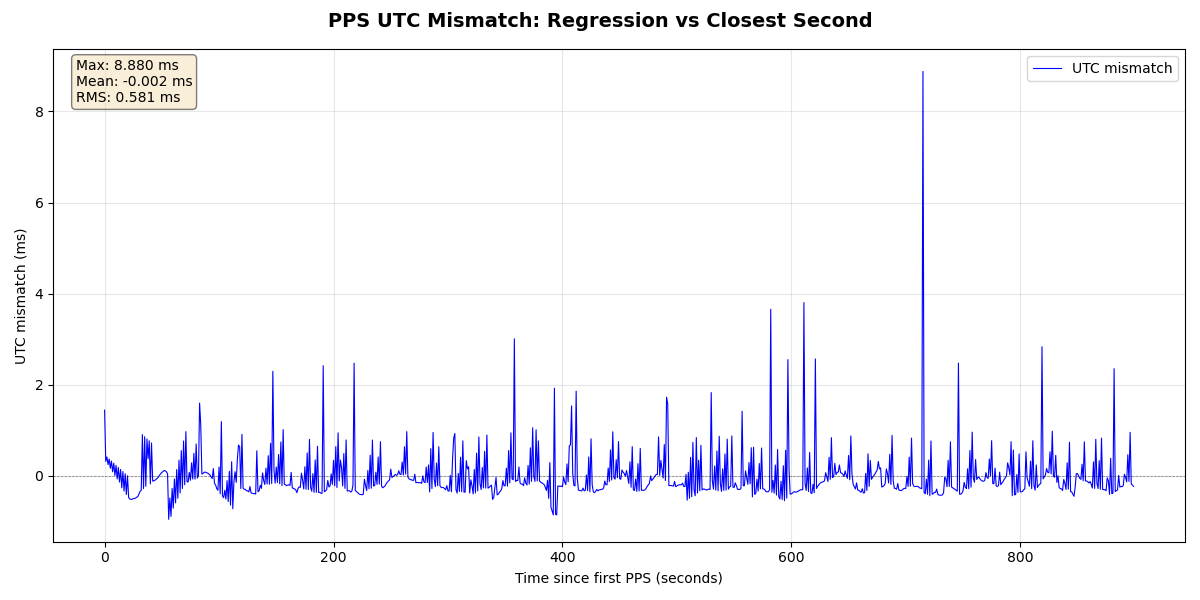}
    \caption{Validation of absolute UTC timestamp accuracy. We show the residual between each logged PPS event and the UTC time predicted by the linear regression of MCU microsecond timestamps to UTC PPS timestamps, for a representative data segment. Residuals are consistently below 10~ms, confirming that the OWL-v2026 produces wave data with better than 10~ms absolute UTC accuracy. This is sufficient for direct cross-correlation analysis between multiple OWL-v2026 units.}
    \label{fig:pps}
\end{figure}

This approach produces high-frequency UTC-timestamped wave data with typically less than 10~ms absolute accuracy, as validated in Figure~\ref{fig:pps}. The decoder can be used as a command-line tool or imported as a Python library.

The decoded data are ready for user-defined post-processing. Wave retrieval from IMU data typically requires applying a complementary filter (e.g., Madgwick or Mahony) or a Kalman filter to fuse the accelerometer and gyroscope channels; implementations are freely available in open-source packages\footnote{An open-source wrapper providing Kalman filter implementation compatible with the OWL-v2026 decoder output is available at \url{https://github.com/gauteh/ahrs-fusion}.} and integrate directly with the decoder output. Beyond the default sensor configuration, the modular architecture also makes the OWL-v2026 readily extensible: one could, for example, substitute a multi-band differential receiver for the GNSS module to obtain absolute heading, or add geophones \citep{voermans2023estimating} or pressure sensors \citep{marchenko2021laboratory} as additional measurement modalities without altering the core firmware architecture.

\section{Test and Validation}
\label{sec:validation}

The OWL-v2026 relies on well-established sensors whose individual performance has been characterized extensively in prior work \citep{sutherland2016observations, rabault2016measurements, rabault2017measurements, rabault2020open, rabault2022openmetbuoy, rabault2024position, muller2025distributed}; we therefore focus our validation on system-level behaviour rather than on sensor characterization. The ISM330DHCX — the same IMU used in the OMB \citep{rabault2022openmetbuoy} — has an accelerometer noise spectral density of $70~\mu\text{g}/\sqrt{\text{Hz}}$ in high-performance mode, corresponding to a detection limit of approximately 0.5~cm wave amplitude at 16~s period in low-noise environments such as sea ice; in rougher open-ocean conditions, accuracy is governed primarily by the post-processing algorithm rather than by sensor noise. The SAM-M10Q GNSS receiver has positioning and Doppler velocity characteristics comparable to those used in other wave-observing systems \citep{rabault2022openmetbuoy, thomson2023development, kodaira2023affordable}, and we refer the reader to those works for detailed accuracy characterizations across sea states and processing approaches. In the following, we report three system-level validation results: logging reliability, power consumption, and absolute UTC timestamp accuracy.

\textbf{Logging reliability and frequency.} We tested continuous logging at 208~Hz (compatible with most SD cards) and 416~Hz (requiring a higher-speed SD card). More than 10 days of continuous testing confirmed no data losses at either frequency, validating that the multi-level buffering scheme successfully absorbs SD card write latency.

\textbf{Power consumption and deployment autonomy.} The dominant power consumer is the SD card. Using a SanDisk Ultra 256~GB card, total logger current draw is approximately 80~mA at 3.3~V. With three SAFT LSH20 D-cell lithium batteries (13~Ah each at 3.6~V, connected in parallel), this gives approximately 20 days of continuous logging. For reference, operating the logger without SD card writes draws only approximately 15~mA, confirming that SD card power consumption dominates. Different SD card models differ markedly in power draw, and testing card choices is advisable before extended deployments.

\textbf{UTC timestamp accuracy.} The PPS-based linear regression approach yields absolute UTC timestamps with typical residuals below 10~ms, as shown in Figure~\ref{fig:pps}. This accuracy is sufficient for direct cross-correlation analysis between multiple OWL-v2026 units.

\section{Conclusion}
\label{sec:conclusion}

We present the OWL-v2026, an open-source, low-cost, easy-to-build, high-performance wave data logger built entirely from off-the-shelf components from the maker community. Its design requires only through-hole soldering for assembly, making it accessible to groups without specialized electronics expertise, and totals approximately 220~USD per unit.

The OWL-v2026 achieves a 10- to 40-fold increase in IMU sampling frequency relative to previous wave loggers that operated at 10--20~Hz \citep{rabault2017measurements}, at approximately one-tenth the cost. Sampling at 208 or 416~Hz resolves rapid wave dynamics that slower loggers cannot capture, including ice floe collisions and wave breaking events. This high sampling rate is complemented by PPS-based UTC synchronization, which delivers better than 10~ms absolute accuracy and enables direct cross-correlation analysis between simultaneously deployed units — a type of analysis that previously required highly specialized setups.

Beyond individual deployments, the low cost opens a qualitatively new mode of field measurement. Arrays of tens to hundreds of OWL-v2026 units could cover areas spanning several wavelengths, enabling direct in-situ analysis of two-dimensional wave propagation and attenuation in ways previously limited to laboratory wave tanks. The OWL-v2026 is also well suited to controlled laboratory experiments when paired with a GNSS repeater to provide indoor satellite signal.

We plan to use the OWL-v2026 in field campaigns investigating wave propagation and floe-floe collisions in sea ice, and we expect its impact on the community to follow a path similar to that of the OpenMetBuoy (OMB) \citep{rabault2022openmetbuoy} and Small FriendlY buoy (SFY) \citep{hope2025sfy}, which substantially expanded the availability of telemetry-enabled wave observations by lowering the cost and barrier to entry. In addition, the OWL-v2026 can be easily extended with virtually any additional sensor connectable via I$^{2}$C, SPI, serial, analog, or digital interface to the OpenLog Artemis board. As the firmware is modular and clearly structured, adding support for new sensors is straightforward, and the existing code provides a clear illustration of how ISR data acquisition and buffering should be implemented.

\section*{Acknowledgements}

We used LLM-powered tools to improve language quality and correctness. All scientific content is the result of the work of the authors, and the authors checked and quality-controlled all LLM-produced language edits. Agentic coding tools were used for the implementation of the logger firmware and the binary data decoder, with significant parts of the code generated by agentic AI tools under author supervision. The authors supervised the development, quality-controlled all programs, performed real-world testing and validation, and validated their outputs.

\appendix

\section{Open Source Release}
\label{app:release}

All code (logger firmware and decoder), design files (PCB schematics and layout), and assembly instructions are available at: \url{https://github.com/jerabaul29/2026_ola_logger_imu_gps}. Questions and support requests can be submitted via the GitHub issue tracker; we will provide reasonable support there.

\section{Lessons Learned from Implementation}
\label{app:lessons}

We document here a set of practical lessons learned during the development and testing of the OWL-v2026, useful for groups developing similar embedded wave logger systems.

\subsection{Hardware design}
\label{app:hardware}

Reliable field electronics require deliberate choices at the hardware level. The following points, learned during the OWL-v2026 development, are of general relevance to similar embedded logger designs.

\begin{itemize}
    \item \textbf{Robust power supply.} Use lithium batteries rated for cold-temperature operation, paired with a step-up/step-down regulator that maintains stable voltage across the full discharge range and at low temperatures.
    \item \textbf{Choose well-supported chips.} Prefer components with mature, well-maintained software libraries to reduce firmware development effort and improve reliability.
    \item \textbf{Avoid bare RF traces.} Use breakout boards from established manufacturers to avoid RF trace design and impedance matching, which are technically demanding and process-dependent.
    \item \textbf{Through-hole only assembly.} Limiting the design to through-hole components at 2.54~mm pitch enables hand assembly without specialized equipment.
    \item \textbf{Master PCB for series assembly.} A master PCB significantly simplifies assembly, reduces errors, and improves reproducibility when building multiple units.
    \item \textbf{Field robustness.} Work in ESD-safe conditions. Include desiccant bags in sealed enclosures. Apply conformal coating to boards where possible. Use IP68 enclosures and supplement with marine-grade silicone for double protection against water ingress.
\end{itemize}

\subsection{Firmware design}
\label{app:firmware}

High-frequency, reliable embedded logging is non-trivial. The following firmware-level lessons were the hardest to learn and are the most likely to be useful to groups developing similar systems.

\begin{itemize}
    \item \textbf{Watchdog timer.} Always use a hardware watchdog timer. Field conditions — vibration, cold, moisture — cause unexpected hardware and software failures; a watchdog ensures automatic recovery.
    \item \textbf{Interrupt-driven and asynchronous architecture.} Reliable high-frequency logging requires interrupt-driven data acquisition paired with non-blocking SD card writes. This introduces concurrency; all shared data structures must be protected using critical sections to prevent race conditions.
    \item \textbf{SD card limitations.} SD cards are poorly suited to the small, latency-sensitive writes imposed by high-frequency logging from a resource-constrained MCU. Modern SD cards are optimized for large sequential transfers, not low-latency small writes. Achieving reliable rates around $\mathcal{O}$(10~kB/s) and higher requires careful buffering, tuning, and card selection. Going significantly beyond this would require a different architecture, such as an external RAM buffer combined with a faster SD interface (e.g., SDIO).
    \item \textbf{Fail-safe error handling.} Implement fault detection at every critical stage: startup, sensor initialization, SD card writes, and the main logging loop. This is essential for reliable unattended field operation.
\end{itemize}

\bibliographystyle{plainnat}

\bibliography{references.bib}

\end{document}